\documentclass[12pt]{article}
\title{Final state interactions: from strangeness to beauty\footnote{
Invited plenary talk at the Chicago Conference on Kaon 
Physics (Kaon'99), June 21-26, 1999, Chicago, IL.}
}

\author{Alexey A. Petrov\\
Johns Hopkins University\\
Baltimore, MD 21218}

\begin{document}
\date{}
\maketitle

\begin{abstract}
I give a brief review of final state 
interactions in meson decays. I describe possible 
effects of FSI in 
$K$, $D$ and $B$ systems, paying particular attention 
to the description of the heavy meson decays.
Available theoretical methods for dealing with 
the effects of FSI are discussed.
\end{abstract}

\section{Motivation}

Final state interactions (FSI) play an important role in meson decays.
The presence of FSI forces one to consider several coupled channels, so
their net effect might be significant, especially if one is 
interested in rare decays. This obvious observation, of course, 
does not exhaust the list of the motivations for better understanding 
of FSI. Many important observables that are sensitive to New Physics 
could also receive contributions from the final state rescatterings.
An excellent example is provided by the $T$-violating
lepton polarizations in $K$ decays (such as $K \to \pi l \nu$ and
$K \to \gamma l \nu$) that are not only sensitive to New Physics
but could also be induced by the electromagnetic FSI.
However, the most phenomenologically important effect of FSI is in 
the decays of $B$ and
$D$ mesons used for studies of direct $CP$-violation, 
where one compares the rates of a $B$ or $D$ meson
decay with the charged conjugated process~\cite{Gronau}. 
The corresponding asymmetries, in order to be non-zero, require 
two different final states produced by different weak amplitudes
which can go into each other by a strong interaction rescattering
and therefore depend on both weak CKM phase and strong
rescattering phase provided by the FSI.
Thus, FSI directly affect the asymmetries and their size can be 
interpreted in terms of fundamental parameters {\it only} if these 
FSI phases are calculable. In all of these examples FSI complicates 
the interpretation of experimental observables in terms of
fundamental parameters~\cite{fknp,dgp}. In this talk I review the progress 
in understanding of FSI in meson decays.

The difference of the physical picture at the energy scales 
relevant to $K$, $D$ and $B$ decays calls 
for a specific descriptions for each class of decays. 
For instance, the relevant energy scale in $K$ decays
is $m_K \ll 1$~GeV. With such a low energy release only 
a few final state channels are available. This significantly 
simplifies the theoretical understanding of FSI in kaon decays. 
In addition, chiral symmetry can also be employed to assist
the theoretical description of FSI in $K$ decays.
In $D$ decays, the relevant scale is $m_D \sim 1$~GeV. This 
region is populated by the light quark resonances, so one might expect 
their significant influence on the decay rates
and $CP$-violating asymmetries. No model-independent 
description of FSI is available, but it is hinted at experimentally 
that the number of available channels is still limited, allowing for a
modeling of the relevant QCD dynamics.
Finally, in $B$ decays, where the relevant energy scale
$m_B \gg 1$~GeV is well above the resonance region, the heavy quark 
limit might prove useful.

\section{Some formal aspects of FSI}
Final state interactions in $A \to f$ arise as a consequence of 
the unitarity of the ${\cal S}$-matrix,
${\cal S}^\dagger {\cal S} = 1$, and involve the rescattering of
physical particles in the final state.
The ${\cal T}$-matrix, ${\cal T} =  i \left (1-{\cal S} \right)$,
obeys the optical theorem:
\begin{equation}
{\cal D}isc~{\cal T}_{A \rightarrow f} \equiv {1 \over 2i}
\left[ \langle f | {\cal T} | A \rangle -
\langle f | {\cal T}^\dagger | A \rangle \right]
= {1 \over 2} \sum_{i} \langle f | {\cal T}^\dagger | i \rangle
\langle i | {\cal T} | A \rangle \ \ ,
\label{unit}
\end{equation}
where ${\cal D}isc$ denotes the discontinuity across physical cut.
Using $CPT$ in the form
$
\langle \bar f | {\cal T} | \bar A \rangle^* =
\langle \bar A | {\cal T^\dagger} | \bar f \rangle =
\langle f | {\cal T^\dagger} | A \rangle,
$
this can be transformed into 
\begin{equation}
\label{opt}
\langle \bar f | {\cal T} | \bar A \rangle^* =
\sum_{i}
\langle f | {\cal S}^\dagger | i \rangle
\langle i | {\cal T} | A \rangle \;\;.
\end{equation}
Here, the states $| i \rangle$ represent all possible final states
(including $| f \rangle$)
which can be reached from the state  $| A \rangle$
by the weak transition matrix ${\cal T}$. The
right hand side of  Eq.~(\ref{opt})
can then be viewed as a weak decay of $| A \rangle$
into $| i \rangle$ followed by a strong rescattering
of $| i \rangle$ into $| f\rangle$. Thus, we
identify $\langle f | {\cal S}^\dagger | i \rangle$
as a FSI rescattering of particles.
Notice that if $| i \rangle$ is an eigenstate of ${\cal S}$
with a phase $e^{2i\delta}$, we have
\begin{equation}
\label{triv}
\langle \bar i| {\cal T} | \bar A \rangle^* =
e^{-2i\delta_i}\langle i | {\cal T} | A \rangle \;\;.
\end{equation}
which implies equal rates for the charge conjugated
decays\footnote{This fact will be important in the studies of
$CP$-violating asymmetries as no $CP$ asymmetry is generated in 
this case.}. Also
\begin{equation}
\langle \bar i | {\cal T} | \bar B \rangle = e^{i\delta} T_i
\langle i | {\cal T} | A \rangle = e^{i\delta} T_i^*
\label{watson}
\end{equation}
The matrix elements $T_i$ are assumed to be
the ``bare'' decay amplitudes and have
no rescattering phases. This implies that these transition
matrix elements between charge conjugated states
are just the complex conjugated ones of each other.
Eq.~(\ref{watson}) is known as Watson's theorem
\cite{watson52}. Note that the problem of unambiguous
separation of ``true bare'' amplitudes from the ``true FSI'' 
ones (known as Omn\'es problem) was solved only for a
limited number of cases.
\subsection{$K$ decays}
The low scale associated with $K$ decays suggests an effective 
theory approach of integrating out heavy particles and making
use of chiral symmetry of QCD. This theory has been known for 
a number of years as chiral perturbation theory ($\chi$PT),
which makes use of the fact that kaons and pions are the Goldstone
bosons of chiral $SU(3)_L \times SU(3)_R$ broken down to $SU(3)_V$,
and are the only relevant degrees of freedom at this energy.  
$\chi$PT allows for a consistent description of the strong and 
electromagnetic FSI in kaon system. 

The discussion of strong FSI is naturally included in the
$\chi$PT calculations of kaon decays processes at one or more 
loops~\cite{SavBij}.
In addition, kaon system is rather unique for its sensitivity to 
the electromagnetic final state interaction effects. Normally, 
one expects this class of corrections to be negligibly 
small. However, in some cases they are still very important. 
For instance, it is known that in non-leptonic $K$ decays the
$\Delta I = 1/2$ isospin amplitude is enhanced compared to the
$\Delta I = 3/2$ amplitude by approximately a factor of 22.
Since electromagnetism does not respect isospin symmetry, one
might expect that electromagnetic FSI might contribute to the
$\Delta I = 3/2$ amplitude at the level
of $22/137\sim 20\%$! Of course, some cancellations
might actually lower the impact of this class of FSI~\cite{DGG}.

There is a separate class of observables that is directly affected 
by electromagnetic FSI. It includes the $T$-violating transverse 
lepton polarizations in the decays $K \to \pi l \nu$ and 
$K \to l \nu \gamma$
\begin{equation}
P^{\perp}_l = \frac{\vec{s}_l\cdot(\vec{p}_i\times\vec{p}_l)}
{|\vec{p}_i\times\vec{p}_l|},
\end{equation}
where $i=\gamma,\pi$. Observation of these polarizations 
in the currently running experiments implies
an effect induced by New Physics. 

A number of parameters of various extensions of the Standard Model
can be constrained via these measurements~\cite{Wu}. 
It is, however, important to realize that the polarizations
as high as $10^{-3}(10^{-6})$ could be generated by the electromagnetic
rescattering of the final state lepton and pion or due to
other intermediate states. These corrections have been estimated for
a number of experimentally interesting final states~\cite{Zhit}.
\subsection{$D$ decays}
The relatively low mass of the charm quark puts the $D$ mesons
in the region populated by the higher excitations of the
light quark resonances. It is therefore natural to assume
that the final state resacttering is dominated by the intermediate 
resonance states~\cite{buccella}. Unfortunately, no model-independent
description exists at this point. Yet, the wealth of experimental
results allows for the introduction of testable models of FSI~\cite{Ros99}.
These models are very important in the studies of direct 
$CP$-violating asymmetries
\begin{equation}\label{cpvas}
A_{CP} = \frac{\Gamma(D \to f) - \Gamma(\bar D \to \bar f)}
{\Gamma(D \to f) + \Gamma(\bar D \to \bar f)} \sim \sin \theta_w
\sin \delta_s,
\end{equation}
which explicitly depend on the values of both weak ($\theta_w$)
and strong ($\delta_s$) phases. In most models of FSI in $D$ decay,
the phase $\delta_s$ is generated by the width of the 
nearby resonance and by calculating the imaginary part of 
loop integral with the final state particles coupled to the 
nearby resonance.

It is important to realize that the large final state interactions 
and the presence of the nearby resonances in the $D$ system has an
immediate impact on the $D-\bar D$ mixing parameters. It is well
known that the short distance contribution to $\Delta m_D$ and
$\Delta \Gamma$ is very small, of the order of $10^{-18}$~GeV.
Nearby resonances can enhance them by one or two orders of
magnitude~\cite{GoP}. In addition, they provide a source for 
quark-hadron duality violations, as they populate the gap 
between the QCD scale and the scale set by the mass of the heavy 
quark normally required for the application of heavy quark expansions.
\subsection{$B$ decays}
In the $B$ system, where the density of the available resonances 
is large due to the increased energy, a different 
approach must be employed. One can use the fact that the 
$b-$quark mass is large compared to the QCD scale and
investigate the behavior of final state phases 
in the $m_b \to \infty$ limit.

Significant energy release in $B$ decays allows  
the studies of inclusive quantities, for instance inclusive 
$CP$-violating asymmetries of the form of Eq.~(\ref{cpvas}).
There, one can use duality arguments to calculate final state 
phases for charmless $B$ decays using perturbative QCD~\cite{bss}. 
Indeed, $b \to c \bar c s$ process, with subsequent final state 
rescattering of the two charmed quarks into the final state (penguin 
diagram) does the job, as for the energy release of the order 
$m_b > 2m_c$ available in $b$ decay, the rescattered $c$-quarks 
can go on-shell generating a CP conserving phase and thus 
${\cal A}_{CP}^{dir}$, which is usually quite small for the 
experimentally feasible decays, ${\cal O}(1\%)$.
It is believed that larger asymmetries can be obtained in exclusive
decays. However, a simple picture is lost because of the 
absence of the duality argument. 

It is known that scattering of high energy particles may be divided 
into `soft' and `hard' parts.  Soft scattering occurs primarily 
in the forward direction with limited transverse momentum 
distribution which falls exponentially with a scale of order 
$0.5$~GeV. At higher transverse momentum one encounters the 
region of hard scattering, which can be described by 
perturbative QCD. In exclusive $B$ decay 
one faces the difficulty of separating the two.
It might prove useful to employ unitarity in trying to 
describe FSI in exclusive $B$ decays.

It is easy to investigate first the {\it elastic} channel.
The inelastic channels have to share a similar asymptotic
behavior in the heavy quark limit due to the unitarity of the 
${\cal S}$-matrix. The choice of elastic channel 
is convenient because of the optical theorem which connects the 
forward (imaginary) invariant amplitude 
${\cal M}$ to the total cross section,
\begin{equation}
{\cal I}m~{\cal M}_{f\to f} (s, ~t = 0) = 2 k
\sqrt{s} \sigma_{f \to {\rm all}} \sim s \sigma_{f \to {\rm all}} \ \ ,
\label{opti}
\end{equation}
where $s$ and $t$ are the usual Mandelstam variables. The asymptotic 
total cross sections are known experimentally to rise slowly 
with energy and can be parameterized by the form \cite{pdg},
$
\sigma (s) = X \left({s/s_0}\right)^{0.08}
+ Y \left({s/s_0}\right)^{-0.56},
$
where $s_0 = {\cal O}(1)$~GeV is a typical hadronic scale.
Considering only the imaginary part of the amplitude, and building in 
the known exponential fall-off of the elastic cross section in $t$  
($t<0$)~\cite{jc} by writing
\begin{equation}
i{\cal I}m~{\cal M}_{f\to f} (s,t) \simeq i \beta_0 \left( {s \over s_0}
\right)^{1.08} e^{bt} \ \ ,
\label{fall}
\end{equation}
one can calculate its contribution to the
unitarity relation for a final state $f = ab$ with kinematics
$p_a' +  p_b' = p_a +  p_b$ and $s = (p_a + p_b )^2$:
\begin{eqnarray}
{\cal D}isc~{\cal M}_{B \to f} &=&
{-i \over 8 \pi^2} \int {d^3p_a' \over 2E_a'}
{d^3p_b' \over 2E_b'}
\delta^{(4)} (p_B - p_a' - p_b') 
{\cal I}m~{\cal M}_{f\to f} (s,t)
{\cal M}_{B \rightarrow f} \nonumber \\
&=& - {1\over 16\pi} {i\beta_0 \over s_0 b}\left( {m_B^2 \over s_0} 
\right)^{0.08} {\cal M}_{B \rightarrow f} \ \ ,
\label{mess}
\end{eqnarray}
where $t = (p_a - p_a')^2 \simeq -s(1 - \cos\theta)/2$,
and $s = m_B^2$.

One can refine the argument further, since
the phenomenology of high energy
scattering is well accounted for by the Regge theory~\cite{jc}.
In the Regge model, scattering amplitudes are described by the 
exchanges of Regge trajectories (families of particles of differing 
spin) with the leading contribution given by the Pomeron
exchange. Calculating the Pomeron contribution to the
elastic final state rescattering in $B \to \pi \pi$
one finds \cite{dgps}
\begin{equation}
{\cal D}isc~{\cal M}_{B \to \pi\pi}|_{\rm Pomeron} = -i\epsilon
{\cal M}_{B \to \pi\pi}, ~~~~~
\epsilon \simeq 0.21 \ \ .
\label{despite}
\end{equation}
It is important that the Pomeron-exchange amplitude is seen to be almost 
purely imaginary. However, of chief significance is the
identified weak dependence of $\epsilon$ on $m_B$  -- the
$(m_B^2)^{0.08}$ factor in the numerator is attenuated by the
$\ln(m_B^2/s_0)$ dependence in the effective value of $b$.

The analysis of the elastic channel suggests that, at high energies,  
FSI phases are {\it mainly generated by inelastic effects}, which
follows from the fact that
the high energy cross section is mostly inelastic. This also
follows from the fact that the Pomeron elastic  
amplitude is almost purely imaginary.  Since the study of
elastic rescattering has yielded a ${\cal T}$-matrix element ${\cal  
T}_{ab
\to ab} = 2 i \epsilon$, i.e. ${\cal S}_{ab \to ab} = 1- 2
\epsilon$, and since the constraint of unitarity of
the ${\cal S}$-matrix
implies that the
off-diagonal elements are ${\cal O}(\sqrt{\epsilon})$,
with $\epsilon$
approximately ${\cal O}(m_B^0)$ in powers of $m_B$ and numerically
$\epsilon < 1$, then the inelastic amplitude must also be ${\cal
O}(m_B^0)$ with $\sqrt{\epsilon} > \epsilon$.
Similar conclusions follow from the consideration of 
the final state unitarity relations. This complements the old 
Bjorken picture of heavy meson decay (the dominance of the
matrix element by the formation of the 
small hadronic configuration which grows into the final state
pion ``far away'' from the point it was produced and does not 
interact with the soft gluon fields present in the decay, see 
also~\cite{DP97} for the discussion) by allowing 
for the rescattering of multiparticle states, production of
whose is favorable in the $m_b \to \infty$ limit, into the
two body final state.
Analysis of the final-state unitarity relations in their general 
form is complicated due to the many contributing intermediate 
states, but we can illustrate the systematics of inelastic 
scattering in a two-channel model. It involves a two-body final 
state $f_1$ undergoing elastic scattering and a final state $f_2$ 
which represents `everything else'. As before, the elastic amplitude 
is purely {\it imaginary}, which dictates the following 
{\it one-parameter} form for the $S$ matrix
\begin{equation}
 S = \left( \begin{array}{cc} 
\cos 2 \theta & i \sin 2 \theta \\
              i \sin 2 \theta & \cos 2 \theta
\end{array} \right)  \ ,\qquad \qquad 
 T = \left( \begin{array}{cc}
2 i \sin^2 \theta &  \sin 2 \theta \\
               \sin 2 \theta & 2 i \sin^2 \theta 
\end{array} \right)  \ \ ,
\label{matr1}
\end{equation}
where we identify $\sin^2 \theta \equiv \epsilon$. The unitarity 
relations become
\begin{eqnarray}
{\cal D}isc ~{\cal{M}}_{B \to f_1} &=& 
- i \sin^2 \theta {\cal{M}}_{B \to f_1} +
\frac{1}{2} \sin 2 \theta {\cal{M}}_{B \to f_2} \ \ ,\nonumber \\
{\cal D}isc~ {\cal {M}}_{B \to f_2} &=& \frac{1}{2} \sin 2 \theta 
{\cal{M}}_{B \to f_1} - i \sin^2 \theta {\cal{M}}_{B \to f_2} \ \ 
\label{big}
\end{eqnarray}
Denoting ${\cal{M}}_1^0$ and ${\cal{M}}_2^0$ 
to be the decay amplitudes in the limit $\theta \to 0$,
an exact solution to Eq.~(\ref{big}) is given by
\begin{equation}
{\cal{M}}_{B \to f_1} = \cos \theta {\cal{M}}_1^0 + i \sin \theta
{\cal{M}}_2^0  \ , \qquad 
{\cal{M}}_{B \to f_2} = \cos \theta {\cal{M}}_2^0 + i \sin \theta
{\cal{M}}_1^0 \ \ .
\label{soln}
\end{equation}
Thus, the phase is given by the inelastic scattering with a 
result of order 
\begin{equation}
{\cal I}m~ {\cal{M}}_{B \to f}/{\cal R}e~ {\cal{M}}_{B \to f} 
\sim 
\sqrt{ \epsilon}~ \left({\cal{M}}_2^0/{\cal{M}}_1^0\right) \ \ .
\end{equation}
Clearly, for physical $B$ decay, we no longer 
have a simple one-parameter ${\cal S}$ matrix, and,
with many channels, cancellations or enhancements are 
possible for the sum of many contributions.
However, the main feature of the above result is expected 
to remain: inelastic channels cannot vanish and provide the 
FSI phase which is systematically of order $\sqrt{\epsilon}$ 
and thus does not vanish in the large $m_B$ limit. 

A contrasting point of view is taken in a recent 
calculation~\cite{bbns}. The argument is 
based on the {\it perturbative} factorization of currents 
(i.e. the absence of infrared singularities) in the matrix 
elements of four quark operators in the Bjorken setup. 
It is claimed that the leading contribution is given by the
naive factorization result with non-leading corrections
suppressed by $\alpha_s$ or $1/m_b$~(see, however,~\cite{IsLS}). 
However, the role of multihadron intermediate states is not yet 
clear in this approach. Moreover, even accepting the result 
of~\cite{bbns}, it would be premature to claim that the 
theory of exclusive $B$ decays to light mesons is free of 
hadronic uncertainties. In fact, many important long distance 
final state rescattering effects involve exchange of global 
quantum numbers, such as charge or strangeness, and thus are 
suppressed by $\approx 1/m_B$. These were shown to be important 
and can be quite large~\cite{fknp,bkp}. This is easy to see in the 
Regge description of FSI where this exchange is mediated by the 
$\rho$ and/or higher lying trajectories. This fact raises a 
question whether the scale $m_b\approx 5$~GeV is large enough 
for the asymptotic limit to set.

{\it (i) Bounds on the FSI Corrections}.
In view of the large theoretical uncertainties~\cite{rosnergronau} 
involved in the calculation of the FSI contributions, it would 
be extremely useful to find a phenomenological method by which 
to bound the magnitude of the FSI contribution. The observation 
of a larger asymmetry would then be a signal for New Physics. 
Here the application of  flavor $SU(3)$ flavor symmetry provides 
powerful methods to obtain a direct upper bound on the FSI 
contribution~\cite{bkp}. 
The simplest example involves bounding FSI in $B \to \pi K$ decays 
using $B^\pm \to K^\pm K$ transitions~\cite{fknp}.

{\it (ii) Direct Observation}. Another interesting way of studying 
FSI involves rare weak decays for which the direct amplitude 
$A(B \rightarrow f)$ is suppressed compared to 
$A(B \rightarrow i)$. They offer a tantalizing possibility of the 
{\it direct observation} of the effects of FSI. 

One of the possibilities involves dynamically suppressed decays which 
proceed via weak annihilation diagrams. It has been argued that
final state interactions, if large enough, can modify the decay 
amplitudes, violating the expected hierarchy of amplitudes. 
For instance, it was shown~\cite{bgr}
that the rescattering from the dominant tree level amplitude
leads to the suppression of the weak annihilation amplitude
by only $\lambda \sim 0.2$ compared to
$f_B/m_B \sim \lambda^2$ obtained from the naive quark
diagram estimate. 

Alternatively, one can study OZI-violating modes, 
i.e. the modes which cannot be realized in 
terms of quark diagrams without annihilation of 
at least one pair of the quarks, like
$\overline{B}^0_d \rightarrow \phi \phi, D^0 \phi$ and 
$J/\psi \phi$. Unitarity implies
that this decay can also proceed via the OZI-allowed weak
transition followed by final state rescattering into the
final state under consideration~\cite{dgps2,GeIsgur}.
In $B$-decays these OZI-allowed
steps involve multiparticle intermediate states and
might provide a source for violation of 
the OZI rule. For instance, the FSI contribution can proceed
via $\overline{B}^0_d \rightarrow \eta^{(\prime)} 
\eta^{(\prime)} \rightarrow \phi \phi$, $\overline{B}^0_d \rightarrow 
D^{\ast 0} \eta^{(\prime)} \rightarrow D^0 \phi$ and $\overline{B}^0_d 
\rightarrow  \psi' \eta^{(\prime)} \rightarrow J/\psi \phi$.
The intermediate state also includes additional 
pions. The weak decay into the intermediate state occurs at tree level, 
through the $(u\overline{u} + d\overline{d})/\sqrt{2}$ component of the 
$\eta^{(\prime)}$ wavefunction, whereas the strong scattering into the 
final state involves the $s\bar{s}$ component.
Hence the possibility of using these decay modes as direct probes of the 
FSI contributions to $B$ decay amplitudes. It is however possible to show that
there exist strong cancellations~\cite{dgps2} among various 
{\it two body} intermediate channels. In the example of 
$\overline{B}^0_d \to \phi \phi$, the cancellation among
$\eta$ and $\eta'$ is almost complete, so the
effect is of the second order in the $SU(3)$-breaking corrections
\begin{equation}
Disc~{\cal M}_{B \to \phi \phi} = O(\delta^2, \Delta^2, 
\delta \Delta) f_\eta F_0 A, ~~\delta = f_{\eta'} - f_\eta,~
\Delta = F_0' - F_0,
\end{equation}
with $A \sim s^{\alpha_0 - 1} e^{i\pi \alpha_0 /2} / 8 b$.
This implies that the OZI-suppressed decays provide an excellent
probe of the multiparticle FSI. Given the very clear signature, these 
decay modes could be probed at the upcoming $B$-factories. 

In conclusion, I reviewed the physics of final state interactions 
in meson decays. One of the main goals of physics of $CP$ violation and 
meson decay is to correctly extract the underlying parameters of 
the fundamental Lagrangian that are responsible for these
phenomena. The understanding of final state interactions 
is very important for the success of this program.

\end{document}